\documentclass[12pt]{article}
\usepackage{amsmath}
\usepackage{graphicx}
\usepackage{amssymb}
\usepackage{amsfonts}
\usepackage{color}
\usepackage{indentfirst}
\usepackage{latexsym}

\def\beq{\begin{eqnarray}}
\def\eeq{\end{eqnarray}}
\def\nn{\nonumber}                

\def\ln{\,\mbox{ln}\,}

\def\al{\alpha}
\def\be{\beta}

\def\de{\delta}

\def\ep{\epsilon}

\def\la{\lambda}
\def\na{\nabla}
\def\pa{\partial}

\def\si{\sigma}
\def\om{\omega}
\def\ph{\varphi}

\def\Ga{\Gamma}

\def\La{\Lambda}

\DeclareMathOperator{\cx}{\square}

\begin{document}

\begin{center}
\renewcommand*{\thefootnote}{\fnsymbol{footnote}}

{\Large
Nonlocality of quantum matter corrections
\\
and cosmological constant running}
\vskip 6mm
		
{E. V. Gorbar}$^{a,b}$
\hspace{-1mm}\footnote{E-mail address: gorbar@bitp.kiev.ua} and
\ \ {Ilya L. Shapiro}$^{c}$
\hspace{-1mm}\footnote{On leave on absence from Tomsk State Pedagogical
University. E-mail address: ilyashapiro2003@ufjf.br},
\vskip 6mm
		
${a)}$ Physics Faculty, Taras Shevchenko National University of Kyiv,
64/13
\\
Volodymyrska Street, 01601 Kyiv, Ukraine
\vskip 2mm

${b)}$
Bogolyubov Institute for Theoretical Physics, Metrologichna 14-b
\\
Kyiv, 03143, Ukraine
\vskip 2mm

${c)}$ Departamento de F\'{\i}sica, ICE,
Universidade Federal de Juiz de Fora
\\
Juiz de Fora, 36036-900, Minas Gerais, Brazil
\end{center}
\vskip 2mm
\vskip 2mm
	
	
\begin{abstract}

\noindent
Semiclassical loop corrections to the gravitational action include
various terms with zero, two, and four derivatives of the metric as
well as nonlocal form factors for these terms. Contributions to some
of these terms could be confused with others on a specific metric
background or for a particular gauge fixing. We present a critical
analysis of the recent works where the tensor structure and the
number of derivatives in the action of gravity were not properly
taken into account. Taking these relevant aspects into account, we
show that although some contributions owing to the quantum
fluctuations of massive or massless scalar as well as fermion and
vector fields may be attributed to the ``running'' of the cosmological
constant, in reality they correspond to the fourth derivative terms
of the action.
\vskip 3mm
		
\end{abstract}
	
\setcounter{footnote}{0} 
\renewcommand*{\thefootnote}{\arabic{footnote}} 

\section{Introduction}
\label{sec1}

\begin{figure}
\begin{quotation}
 \mbox{\hspace{+2.9cm}}
\includegraphics[width=9.0cm,angle=0]{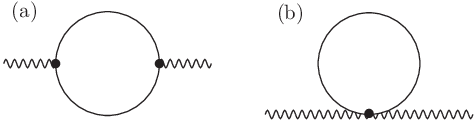}
\caption{{\sl
Two diagrams of the one-loop correction to
$\langle h_{\mu\nu} h_{\al\be} \rangle$ due to a quantum matter
field represented by thick lines. Diagram (a) gives a
contribution to the nonlocal form factor, while diagram (b)
contributes only to the local expression, i.e., divergences.
Both plots are needed to provide a correspondence between
logarithmic divergences and the logarithmic UV limit for the form
factors \cite{apco}.}}
\label{Fig1}
\end{quotation}
\end{figure}

Quantum matter corrections to the effective gravitational action is
one of the main points of interest in semiclassical gravity staring
from the seminal pioneering works \cite{UtDW,StaZel71}. A
characteristic feature of the contributions of massive quantum field
is the presence of more complicated nonlocal form factors in the
vacuum effective action \cite{bavi90,avramidi,apco,fervi}\footnote{
See also subsequent works \cite{CodelloZanusso2013,Omar4D} for
the extension to the total derivative form factor of the Einstein
term.} that boil down to the logarithmic expressions in the UV (at
large energies) \cite{bavi85,bavi90}. The references
\cite{bavi90,avramidi,apco,fervi,Omar4D} employ the heat-kernel
technique providing one of the most direct and unambiguous methods
to derive the mentioned form factors. An alternative equivalent
approach is to use Feynman diagrams on the flat Minkowski spacetime
background $\eta_{\mu\nu}$  \cite{apco,CodelloZanusso2013}.
Decomposing the metric as follows
\beq
g_{\mu\nu} \,=\, \eta_{\mu\nu} +  h_{\mu\nu},
\label{flat}
\eeq
one has to calculate the self-energy diagrams (or the
polarization operator \cite{StaZel71}) for the graviton propagator
$G_{\mu\nu\al\be}=-i\langle h_{\mu\nu} h_{\al\be} \rangle$
shown in Fig.~\ref{Fig1} and then determine the semiclassical
corrections to the gravitational action.

The role of mass for the problem under consideration is crucial.
At low energy (in the IR), in the momentum representation, one can
make an expansion in the ratio $k^2/m^2$ (where  $k^2 = k^\mu k_\mu$
is the square of Euclidean momentum) and arrive at the gravitational
analog of the Appelquist and Carazzone decoupling theorem \cite{AC}.
This decoupling has been found in \cite{apco}, but only for the
$C^2$ and $R^2$ terms, where $C^{\mu\nu\al\be}$ is the Weyl tensor.
In a recent work
\cite{Omar4D}, one can see the decoupling taking place for the
Einstein-Hilbert term, but not for the cosmological constant term.

One of the interesting outputs of calculations \cite{apco} is that
they do not provide a nonlocal form factor for the cosmological constant term.
The reason is that  the hypothetical nonlocal form favor $k_\La(\cx)$
should act on a constant term, giving zero.
However, it was suggested in \cite{apco} and in a slightly different
framework in \cite{DCCrun} that an effective running of the
cosmological term $\rho_\La$ would be possible if we meet, in
the quantum corrections to the gravitational action, the logarithm
form factors in the terms quadratic in the Ricci tensor or the curvature
scalar,
\beq
\int d^4x \,\sqrt{-g}\,
R_{\mu\nu}\,\frac{m^4}{\cx^2}\, R^{\mu\nu}
\,,\qquad
\int d^4x \,\sqrt{-g}\,
R\,\,\frac{m^4}{\cx^2}\, R\,,
\label{nonlocCC}
\eeq
or in a similar term with the Riemann tensor, or in higher order terms
such as $\int F(\cx^{-1}R)$, etc.
To see the reason why terms (\ref{nonlocCC})
are nonlocal analogs of the cosmological constant term
\beq
-\int d^4x \,\sqrt{-g}\rho_\La\,,
\qquad
\rho_\La = \frac{\La}{8\pi G}
\label{rhoCC}
\eeq
(called the CC term, in what follows), one has to consider a local
(similar to the cosmological evolution) scaling of the metric
\beq
g_{\mu\nu} \,\,\,\longrightarrow \,\,\, a^2(\eta) g_{\mu\nu} ,
\label{scaling}
\eeq
where $\eta$ is the conformal time. For $a=const$, terms
(\ref{nonlocCC}) scale exactly like the CC term.
Euristically, the d´Alembert operator in denominator is
traded for two derivatives in $R_{\mu\nu}$ or $R$. For a
time-dependent $a(\eta)$, these nonlocal terms provide typically
a mild deviation from the CC term and this explains the phenomenological
success of the corresponding cosmological models \cite{Magg2}
(see also further references therein). In general, cosmological
models with slowly varying CC term attract a lot of attention and
for this reason, it would be very interesting to derive the nonlocal
terms (\ref{nonlocCC}) explicitly. However, the attempts to do so,
failed so far. In particular, the nonlocal terms that emerge in the
vacuum action from a spontaneous symmetry breaking (SSB)
\cite{Spont} (see also more detailed consideration, including highly
non-trivial derivation of the energy-momentum tensor of vacuum,
in \cite{Tmn-ABL}) produced a negative result. Although there are
many nonlocal terms in the SSB-induced action, the ones of type
(\ref{nonlocCC}) cancel out.

In this situation, it looks very promising that similar terms with
extra logarithm factor
\beq
\mathcal{L}\,=\,
\frac{m^4}{40\pi^2}
\bigg\{
\Big( \frac{1}{\cx}R_{\la\si}\Big)
\ln\Big( \frac{\cx + m^2}{m^2}\Big)
\,\Big( \frac{1}{\cx}R_{\la\si}\Big)
\,-\,\frac18 \Big( \frac{1}{\cx}R\Big)
\ln\Big( \frac{\cx + m^2}{m^2}\Big)
\Big( \frac{1}{\cx}R\Big)
\bigg\}
\label{nonlocCCmass}
\eeq
were found in a recent work \cite{Donoghue_22}.
According to \cite{Donoghue_22}, these expressions were obtained
from the very same diagrams in Fig.~\ref{Fig1} calculated in
\cite{apco,CodelloZanusso2013}
and this fact requires, in view of the above mentioned, an explanation.

An alternative description of quantum matter effects is
the renormalization group running. The physical beta functions can
be derived using the momentum-subtraction scheme of renormalization.
In the IR, these beta functions demonstrate the quadratic decoupling,
however, only for the coefficients of the $C^{2}$, $R^2$
terms \cite{apco}, and the $R$-term \cite{Omar4D}.

It is worth noting that in the case of the minimal subtraction (MS)
renormalization scheme, there is a well-defined beta function for
$\rho_\La$  \cite{Brown,OUP}. An interesting paper \cite{Woodard_08}
entitled ``Cosmology is not a renormalization group flow'' compares
the MS-scheme beta function for $\rho_\La$ with the explicit
calculation of quantum contributions to the vacuum energy on de
Sitter space and finds different results. However, this comparison was
made for the quantum effects of a  \textit{massless} scalar field.
On the other hand, the beta function for $\rho_\La$ vanishes in the
MS or any other scheme of renormalization for all known types of
massless fields. Thus, one has to explain the discrepancy between
the vanishing beta function and the nonzero result on a de Sitter
background.

It turns out that the explanation of the two results concerning the
quantum contributions to the vacuum energy density $\rho_\La$
\cite{Donoghue_22} and \cite{Woodard_08}, is quite similar,
regardless that they were obtained by completely different methods.
In both cases, subtleties in the interpretation of the calculations
are connected with the tensor structure of the gravitational action.
We believe that explaining these results in more detail will be
instructive and present the corresponding explanations in the
rest of this paper.

The work is organized as follows. Section~\ref{sec.2} briefly
reviews the expression for the nonlocal effective action derived
in \cite{apco}. In Sec.~\ref{sec.3}, we discuss quantum matter
corrections in de Sitter space and in Sec.~\ref{sec.4} provide an
in detail analysis of the results of the calculation of Feynman
diagrams in \cite{Donoghue_22}. Finally, we draw our conclusions
in Sec.~\ref{sec.5}.

\section{Form factors for the gravitational action}
\label{sec.2}

Let us start by introducing some standard notions. The general
expression for the action of semiclassical gravity providing
renormalizable semiclassical theory, has the form
\beq
S_{vacuum}\, =\,
S_{EH}\, +\,
S_{HD}
\label{vacuum}
\eeq
and consists of the Einstein-Hilbert term
\beq
S_{EH}\, =\,
-\,\frac{1}{16\pi G}\,\int d^4x\,\sqrt{-g}\,(R + 2\La)
\label{Einstein}
\eeq
and the higher derivative term
\beq
S_{HD}\, =\, \int d^4x\,\sqrt{-g}\,\left\{\,
a_1 C^2 + a_2 R^2 + a_3 E_4 + a_4 \cx R \,\right\}\,,
\label{higher}
\eeq
which includes the square of the Weyl tensor
$C^2=R_{\mu\nu\al\be}^2 - 2R_{\al\be}^2 + 1/3\,R^2$
and the integrand of the Gauss-Bonnet topological invariant
$E_4 = R_{\mu\nu\al\be}R^{\mu\nu\al\be}
-4 \,R_{\al\be}R^{\al\be} + R^2$. Here
$\,a_1$,..., $a_4$, $\,G\,$, and $\,\La\,$ are the parameters
of the semiclassical gravitational action.
As in \cite{apco} and \cite{Donoghue_22}, we consider quantum
matter corrections due to a free {\it massive} scalar field
\beq
S_m\,\,=\,\,\frac12\int d^4x\,\sqrt{-g}\,
\left\{g^{\mu\nu}\pa_\mu\ph\pa_\nu\ph
 - m^2\ph^2 + \xi R\ph^2 \right\}
\label{scalar}
\eeq
with the nonminimal parameter of interaction $\xi$.

The one-loop contribution $\bar{\Ga}_{vac}$
to the gravitational action was obtained in \cite{apco} by
using the diagram approach with the expansion in the curvature
tensor and its covariant derivatives up to the second order in
curvature (and verified employing the heat kernel method).
The result of these calculations has the form
\beq
&&
\bar{\Ga}_{vac}
\,=\,\frac{1}{2(4\pi)^2}\,\int d^4x \,\sqrt{-g}\,
\bigg\{\,\frac{m^4}{2}\,
\Big(\frac{1}{\ep_\mu} \,+\,\frac32\Big)
\,+\,\tilde{\xi} m^2R\,\Big[\frac{1}{\ep_\mu}\,+\,1\Big]
\nn
\\
&&
\qquad
\,\,
+\,\frac12\,C_{\mu\nu\al\be} \,\Big[\frac{1}{60\ep_\mu}\,+\,k_W(a)
\,\Big] C^{\mu\nu\al\be}
\,+\,R \,\Big[\,
\frac{1}{2\ep_\mu}\,\tilde{\xi}^2 \,+ \,k_R(a)\,\Big]\,R\,\bigg\}\,,
\qquad
\label{final}
\eeq
with nonlocal form factors
\beq
&&
k_W(a)\, = \,\frac{8A}{15\,a^4}
\,+\,\frac{2}{45\,a^2}\,+\,\frac{1}{150}\,,
\nn
\\
&&
k_R(a)\, =
\,
\Big(A + \frac{1}{18}\Big)\tilde{\xi}^2
+ \Big(\frac{2A}{3a^2} - \frac{A}{6}\Big)\tilde{\xi}
+ \frac{A}{9a^4}-\frac{A}{18a^2}+\frac{A}{144}
+ \frac{1}{108\,a^2}
- \frac{7}{2160}\,.
\qquad
\label{C}
\eeq
In these formulas, we consider Euclidean signature and use
the notations
\beq
&&
\frac{1}{\ep_\mu}\,=\,
\frac{1}{2-w}
+ \ln \Big(\frac{4\pi \mu^2}{m^2}\Big)\,,
\qquad
\tilde{\xi} = \xi-\frac16
,\qquad
a^2=\frac{4\cx}{\cx-4m^2}\,,
\label{xi}
\nn
\\
&&
A =-\frac12\int_0^1 d\al\ln\Big[1+\al(1-\al)u\Big]
=1-\frac{1}{a}\ln \frac{1+a/2}{1-a/2}\,.
\label{A}
\eeq
Two things are worth mentioning here.

\textit{i)} The one-loop contribution (\ref{final}) is complete in
the $\mathcal{O}(R^2_{...})$ approximation.

\textit{ii)} In the massless case, there is no $m^4$-type counterterm
and the $MS$-scheme beta function for $\rho_\La$ vanishes. Let us
note, in passing, that in the massless case there are no corrections
to the CC term also at higher loops. In both massive and massless
cases, there is no nonlocal form factor for the CC term. Only the
local counterterm $\propto m^4$ is present in the massive case.

Results \textit{i)} and \textit{ii)} are in a direct conflict with the
interpretations of  \cite{Woodard_08} and \cite{Donoghue_22}, where
the quantum correction to the CC cosmological constant equation of
state $\om=-1$ in the massless theory and the $m^4$-type
\textit{nonlocal} one-loop correction to the CC were
found, respectively. We clarify an apparent contradiction of the
obtained results in the sections below and show that all results are
consistent taking into account that the loop contributions to the
vacuum action correspond not only to the $\rho_\La$-term but, also,
to other relevant terms in the classical action (\ref{vacuum}).
However, only terms (\ref{higher}) have logarithmic divergences
and, therefore, only these terms gain logarithm ($\cx$-dependent)
form factors from the loop corrections.

\section{IR dependence on $a$ in the massless case}
\label{sec.3}

The logic of \cite{Woodard_08} is based on the comparison of the
renormalization group equation for $\rho_\La$ in \cite{AntMot},
\cite{CC-nova}, and other papers on the renormalization group running
of the CC and the quantum corrections to the vacuum equation of state
obtained in \cite{Onemli-Woodard2004}. The last calculation is based
on the well-defined stochastic quantization of massive scalar field
on a de Sitter background \cite{StarobinskyYakoyama1994}. This
technique enables one to obtain higher-loop perturbative and even
nonperturbative results.\footnote{Compared to the recent attempt to
apply the stochastic formalism to the case of an arbitrary metric 
background
\cite{Stock-GK}, using the de Sitter metric provides significant
advantages. As we will show below, this also means the need for
a special care in the interpretation of results.}
The massless limit in this method is smooth and one should expect
a good fit with other methods of calculations.

The first-order logarithm corrections for the energy density and the
pressure of vacuum, quoted in \cite{Woodard_08}, have the form
\beq
&&
\de \rho_\La \,=\, - \,\frac{\la}{(16 \pi^2)^2}\,
\frac{7H^4}{2}\,\ln(a)
,
\nn
\\
&&
\de \big(\rho_\La + p_\La\big)
\,=\, - \,\frac{\la}{(16 \pi^2)^2}\,\frac{4H^4}{3}\,\ln(a).
\label{Wood-IR}
\eeq
One can identify these terms as two-loop contributions because
they are proportional to the coupling constant $\la$ of the $\ph^4$
interaction. At one loop, it is very well known that the vacuum
contributions do not have such a dependence. On the other hand,
in a usual perturbative treatment, there are  $\ln(a)$-terms but only
in the effective action and \textit{not} in the elements of the
energy-momentum tensor, such as $\rho_{vac}$ and $p_{vac}$.

Let us start the discussion by saying that we do not contest the
correctness of the calculations leading to  (\ref{Wood-IR}). On
the other hand, we believe that these calculations should be
correctly interpreted. It looks misleading to compare them with the
running of the cosmological constant term, described in different
frameworks (e.g., in \cite{nelspan82} or \cite{AntMot}).

First, we can pose a question which may be useful as a starting point.
Since the one-loop contribution to the effective action (\ref{final})
has nor CC-type divergences, neither nonlocal corrections to the CC
term in the massless case,
how one can resolve the contradiction between (\ref{Wood-IR})
and (\ref{final}) in the limit $m \to 0$?

Let us recall that the calculations of \cite{Onemli-Woodard2004}
were done on a de Sitter background, where the terms in action
(\ref{vacuum}) boil down to the expressions
\beq
E_4 = \frac16\,R^2,
\qquad
C^2 = 0,
\quad
\mbox{and}
\quad
R = -12 H^2.
\label{dS}
\eeq
Taking this into account and looking at Eq.~(\ref{Wood-IR}),
we readily see that the logarithmic ``running'' should not be
attributed to the CC term $\rho_\Lambda$, but rather to a linear
combination of $E_4 = 24H^4$ and  $R^2 = 144H^4$.
Indeed, when making the
comparison of the running of the parameters in action
(\ref{final}) with the output of de Sitter calculations, the unique
reasonable attribution is to the running of $a_2$ and $a_3$
in Eq.~(\ref{higher})
and not of the CC term. The reason is (as we mentioned above)
that the constant $\rho_\La$ does not run in massless models
and, also, it is not $\mathcal{O}(H^4)$, while the $E_4$ and
$R^2$-terms are exactly of order $H^4$  
One can note that, in the conformal ($\xi=1/6$) massless case, the
corrections to $\rho$ and $p$ can be easily recovered from the trace
anomaly \cite{duff77,ddi} and the conservation equation, as described
in \cite{fhh} and, more recently, in \cite{RadiAna} (see further
references therein). Unfortunately, our attempt to recover the
coefficients in (\ref{Wood-IR}) from the known one-loop beta
functions did not work well, probably because of the second loop
``contamination''.

Another relevant observation is that, in the higher
derivative model \cite{AntMot}, there are both massless and
massive modes, but it is not clear how to separate them in the IR
and arrive at the meaningful comparison with the IR evolution
described in \cite{Onemli-Woodard2004} in the massless case.

It would be certainly interesting to find an analog of the formulas
(\ref{Wood-IR}), but this is not an easy task. Let us start
from a simpler problem and present the expressions for the
one-loop four-derivative contributions on de Sitter space
for a conformal scalar field. In this case, the trace anomaly
has the form
\beq
&&
\langle T^\mu_{\,\,\mu} \rangle \,=\,\om C^2
+ b E_4
+ c {\Box}R ,
\nn
\\
&&
\om = \frac{1}{120\,(4 \pi)^2},
\quad
b = - \frac{1}{360\,(4 \pi)^2},
\quad
c = \frac{1}{180\,(4 \pi)^2}.
\label{T-II}
\eeq

One can make a comparison between the anomaly and the
anomaly-induced action, which can be presented in the
simplest form using the variables
$\,{\bar g}_{\mu\nu}\,$ and $\,\si = \ln(a)$, where
\beq
{g}_{\mu\nu} = {\bar g}_{\mu\nu}\cdot e^{2\sigma}.
\label{confmet}
\eeq
The standard non-covariant form of the induced action is
\beq
&&
{\bar \Gamma} = \int d^4 x
\sqrt{-{\bar g}} \,\bigg\{
\om \sigma{\bar C}^2 + b\sigma
({\bar E}_4 - \frac23 \,{\bar {\Box}} {\bar R})
+ 2 b\,\sigma{\bar \Delta}_4\sigma
\nn
\\
&&
\qquad \qquad
\,- \,\frac{3c+2b}{36}\,
\big[\bar{R} - 6 (\bar{\na}\si)^2  - 6 \bar{\cx}\si \big]^2
\bigg\}\,.
\label{quantum}
\eeq
In the case of de Sitter background, there is a term
$b\si {\bar E}_4$, which becomes $24 H^4 \log (a)$ on de Sitter
background.

In the cosmological setting, e.g., assuming a conformally flat
FRW metric $g_{\mu\nu} = a^2(\eta) \eta_{\mu\nu}$, one can
combine the conservation law and the relation for the trace, i.e.,
\beq
d\,(\rho_{vac} \,a^3) \, = \, - \,p_{vac}\,d(a^3)
\quad \mbox{and} \quad
\langle T^\mu_{\,\,\mu} \rangle
\,=\,\rho_{vac} - 3p_{vac},
\label{conslaw}
\eeq
to arrive at the one-loop contributions to the energy density and
pressure of the vacuum \cite{fhh,RadiAna},
\beq
{\bar \rho}_{vac}
&=& - 6c\left[\frac{a^{\prime\prime\prime}a^{\prime}}{a^6}
-\frac{1}{2}\left(\frac{a^{\prime\prime}}{a^3}\right)^2
-2\frac{a^{\prime\prime}a^{\prime\, 2}}{a^7}
\right]+6b \left(\frac{a^{\prime}}{a^2}\right)^4 ,
\nonumber
\\
{\bar p}_{vac} &=&
-2c\left[5\frac{a^{\prime\prime\prime}a^{\prime}}{a^6}
- \frac{a^{(IV)}}{a^5}
+\frac{5}{2}\left(\frac{a^{\prime\prime}}{a^3}\right)^2
-8\frac{a^{\prime\prime}a^{\prime\, 2}}{a^7}\right]
+8b \left[3\left(\frac{a^{\prime}}{a^2}\right)^4
-\frac{a^{\prime\prime}a^{\prime\, 2}}{a^7}\right] ,
\label{EOS HD}
\eeq
where the prime stands for the derivative with respect to conformal
time $\eta$. On the de Sitter background, with $H=a'/a^2=const$,
this boils down to
\beq
{\bar \rho}^{dS}_{vac} = (6b+2c)H^4,
\qquad
{\bar p}^{dS}_{vac} =  (8b+2c)H^4.
\label{EOS dS}
\eeq
Compared to the formulas (\ref{Wood-IR}), we can see that the
relations (\ref{EOS dS}) nicely reproduce the general factors
$H^4$. This is explained by the fact that (\ref{Wood-IR}) should
not be interpreted as a correction to the cosmological constant
term, but to the fourth derivative terms. On the other hand, there
are the following serious differences between (\ref{EOS dS}) and
(\ref{Wood-IR}):
\vskip 1mm

\textit{i)} The relations (\ref{EOS dS}) do not depend on the
coupling constant $\la$, which is present in (\ref{Wood-IR}).
This shows that the last formula is a two-loop contribution,
while the former comes from the one-loop contribution.
\vskip 1mm

\textit{ii)} Different from the effective action (\ref{quantum}),
the relations (\ref{EOS dS}) do not have the logarithms of $a$,
which are present in (\ref{Wood-IR}). This is certainly related
to the previous point. At the second loop, usually there are terms
quadratic in the logarithms which should produce $\ln (a)$ in
the ${\bar \rho}^{dS}_{vac}$ and ${\bar p}^{dS}_{vac}$.
However, in the massless theory, these logarithms describe the
fourth derivative terms and not the one of the cosmological
constant term.
\vskip 1mm

\textit{iii)} The formulas (\ref{EOS dS}) correspond to the
conformal scalar, while  (\ref{Wood-IR}) correspond to the
minimal scalar. This difference is not critical and can be
eliminated by assuming the non-local running of the coefficient
of the $R^2$-term in the minimal theory. A pertinent observation
is in order in this respect. The minimal interacting scalar theory
is not renormalizable even at the one-loop level, different from
the conformal scalar theory.  For this reason, the vacuum
effective action in the minimal version of the scalar theory
is plagued by severe non-renormalizable divergences at the
two loop order. This maybe not a problem for stochastic
formalism, on which the derivation of   (\ref{Wood-IR}) was
based. However, this point makes any kind of comparison with
the usual perturbative formalism a completely non-trivial issue.

As we mentioned above,
the expression (\ref{quantum}) represents a local version of the
renormalization group running which, in difference to a global
scaling (as correctly noted in \cite{Woodard_08}), can be
consistently applied to cosmology. However, this action is
well-defined only for a classically conformal theory and mainly
at the one-loop level. On the contrary, the relations (\ref{Wood-IR})
and their higher-order generalizations
\cite{StarobinskyYakoyama1994,Onemli-Woodard2004} can be
used far beyond the framework of (\ref{quantum}), e.g., in the
non-conformal models and at higher loops. However,
the consistent use of these important results is possible only on
the basis of their correct interpretation. We hope that the present
communication will contribute in this direction.

\section{Nonlocal $m^4$-type form factors}
\label{sec.4}

Let us return to the remarkable result of
\cite{Donoghue_22} given by Eq.(\ref{nonlocCCmass})
suggesting an intriguing nonlocal partner for the CC
term in the gravitational action. This seems to be a very good
progress compared to expression (\ref{final}), where one cannot
find any form factor for the $\rho_\La$ term. Therefore, first of
all, we should check whether (\ref{final}) is compatible with
(\ref{nonlocCCmass}) derived from the same diagrams in
Fig.~\ref{Fig1}.

The diagram calculation in \cite{apco} is based on the decomposition
of the polarization operator in different tensor structures and using
expansion (\ref{flat}) to the second order in $h_{\mu\nu}$,
\beq
&&
\int d^4x \sqrt{-g}\,=\,\int d^4x\,h^{\mu\nu}\,
\Big(\,\frac18\,\eta_{\mu\nu}\,\eta_{\al\be}
-\,\frac14\,\de_{\mu\nu\,,\,\al\be}\,\Big)\,h^{\al\be}\,+...\,,
\label{CC}
\\
&&
\int d^4x \sqrt{-g} R\,=\,\int d^4x\,\,\,h^{\mu\nu}\,
\Big[\,\frac14\,\big( \de_{\mu\nu\,,\,\al\be}
-\eta_{\mu\nu}\,\eta_{\al\be}\big)\cx
-\,\frac12\,\eta_{\mu\al}\pa_\nu\pa_\be
\nn
\\
&&
\qquad \,\,
\qquad \,\,
\qquad \,\,
+ \,\frac14\,
(\eta_{\mu\nu}\pa_\al\pa_\be + \eta_{\al\be}\pa_\mu\pa_\nu)
\,\Big]\,h^{\al\be}
\,+...\,,
\label{EH-expand}
\eeq
\beq
&&
\int d^4x \sqrt{-g} R^2=\int d^4x\,\,h^{\mu\nu}
\Big[\,\pa_\mu\pa_\nu \pa_\al \pa_\be
+\eta_{\mu\nu}\eta_{\al\be}\cx^2
\nn
\\
&&
\qquad \,\,
\qquad \,\,
\qquad \,\,
- \,(\eta_{\mu\nu}\pa_\al\pa_\be
+ \eta_{\al\be}\pa_\mu\pa_\nu)\cx
\Big]h^{\al\be}
+..\,,
\eeq
\beq
&&
\int d^4x \sqrt{-g}\,W\,=\,\int d^4x\,h^{\mu\nu}\,
\Big[\frac14\,\Big(\de_{\mu\nu\,,\,\al\be}\,-\,
\frac{1}{3}\,\eta_{\mu\nu}\,\eta_{\al\be}\Big)\cx^2
\,-\, \frac12\,\eta_{\mu\al}\,\pa_\be\pa_\nu\cx
\nn
\\
&&
\qquad \,\,
\qquad \,\,
\qquad \,\,
+ \,\frac{1}{12}\,(\eta_{\mu\nu}\pa_\al\pa_\be
+ \eta_{\al\be}\pa_\mu\pa_\nu)\,\cx
\,+\, \frac16\,\pa_\mu\pa_\nu\pa_\al\pa_\be
\Big]\,h^{\al\be}\,+...\,,
\mbox{\qquad}
\label{W}
\eeq
where dots stand for the lower and higher orders (in $h_{\mu\nu}$)
terms. Here we used the notation $2W = C^2 - E_4$ which simplifies
the expansion without the loss of generality. It is easy to see that
there are five distinct tensor structures. In momentum space,
we have \cite{OUP}
\beq
&&
{\hat T}_1 = \de_{\mu\nu,\al\be}
= \frac12\big( \eta_{\mu\al} \eta_{\nu\be}
+ \eta_{\mu\be} \eta_{\nu\al}\big),
\qquad
{\hat T}_2 = \eta_{\mu\nu} \eta_{\al\be},
\qquad
{\hat T}_3 = \frac{1}{k^2}
\big( \eta_{\mu\nu} k_\al k_\be + \eta_{\al\be} k_\mu k_\nu \big),
\nn
\\
&&
{\hat T}_4\,=\,\frac{1}{4k^2}
\big(\eta_{\mu\al}k_\be k_\nu
+ \eta_{\nu\al}k_\be k_\mu
+ \eta_{\mu\be}k_\al k_\nu
+ \eta_{\nu\be}k_\al k_\mu \big)\,,
\quad
{\hat T}_5\,=\, \frac{1}{k^4}\, k_\al k_\be k_\mu k_\nu .
\mbox{\qquad}
\label{biliH-T}
\eeq
After decomposing the polarization operator in these tensor
structures, one should calculate the loop integrals and arrive
at the expressions equivalent to (\ref{final}).

The calculations in \cite{Donoghue_22}
could be done either using the gauge fixing condition
$\pa_\mu h^\mu_{\,\,\nu}= \frac12 \pa_\nu h$ for the
external metric perturbation\footnote{At least,
this is stated in  \cite{Donoghue_22}.}  (\ref{flat}), or
simply taking into account only the
first two tensor structures ${\hat T}_1$ and ${\hat T}_2$
in (\ref{biliH-T}). In both cases, the two tensor structures
${\hat T}_{1,2}$ in (\ref{biliH-T}) are insufficient to distinguish
different terms in the expansion (\ref{W}).  The unique remaining
criterion is that the term $W$ (or, equivalently, $C^2$) contributes
to the propagator of the traceless mode
$\bar{h}_{\mu\nu}=h_{\mu\nu} - \frac14 h g_{\mu\nu}$ rather
than the propagation of the trace mode $h = h^\mu_{\,\,\mu}$, while
the $R^2$-term contributes to the propagator of the trace mode only.
Both aforementioned procedures (using gauge fixing
for the external field $h_{\mu\nu}$ or taking into account only the
two tensor structures instead of five), enable one to separate the
two terms in (\ref{nonlocCCmass}). This means that the calculations 
of \cite{Donoghue_22} are almost correct, but their correct
interpretation requires taking into account the remaining tensor
structures ${\hat T}_{3,4,5}$ in (\ref{biliH-T}). Only in this way
one could avoid
the mixing of the contributions to the $C^2$ and $R^2$ form
factors with the ones for the CC and Einstein-Hilbert terms.

Thus, the problem is to extract (\ref{nonlocCCmass})
from the full expression (\ref{final}) under the gauge fixing
condition $\pa_\mu h^\mu_{\,\,\nu}= \frac12 \pa_\nu h$, where
we meet only the two tensor structures ${\hat T}_1$  and
${\hat T}_2$.  Let us present the corresponding procedure
in detail.

For both nonlocal form factors $k_W(a)$  and  $k_R(a)$,
logarithm terms are present \textit{only} in the UV, thus, we
have to consider the regime $k^2 \gg m^2$. In this case, by using
the relations
\beq
\frac{4}{a^2} = 1+ \frac{4m^2}{k^2}\,,
\qquad
\ln \frac{1+a/2}{1-a/2}
\,\,\longrightarrow\,\, \ln \Big(\frac{k^2}{m^2}\Big),
\label{interm}
\eeq
we obtain
\beq
&&
k_W \,=\, - \, \Big[
\frac{1}{60} + \frac{1}{6}\,\frac{m^2} {k^2}
+ \frac{1}{2}\,\frac{m^4}{k^4}\Big]\, \ln \Big(\frac{k^2}{m^2}\Big)
\,+\,\dots\,\,,
\nn
\\
&&
k_R \,=\, -\,\bigg[\frac{\tilde{\xi}^2}{2}
+ \Big( \frac{\tilde{\xi}}{3} +  \tilde{\xi}^2\Big)
\,\frac{m^2}{k^2}
\,+\, \Big( \frac{1}{18} + \frac{2\tilde{\xi}}{3} -  \tilde{\xi}^2\Big)
\,\frac{m^4}{k^4}\bigg]\, \ln \Big(\frac{k^2}{m^2}\Big)\,+\,\dots\,\,,
\label{Ops}
\eeq
where dots indicate $\mathcal{O}(m^6/k^6)$ and non-logarithm terms.
One can extract the UV limit of the one-loop contributions
(\ref{final}), present it in the coordinate representation, and use the
Ricci basis, i.e., replace $(1/2)C^2 \to W = R_{\la\si}^2 - (1/3)R^2$.
This procedure gives us the following expression for the effective
Lagrangian (we change the signature to the Lorentz one to have
correspondence with \cite{Donoghue_22}):
\beq
&&
\mathcal{L}^{(UV,2)}_{scal}\,=\,
\frac{1}{(4\pi)^2}\bigg\{\,\frac{m^4}{4}
\Big(\frac{1}{\ep_\mu}+\frac32\Big)
\,+\,
\frac{\tilde{\xi} m^2}{2}\,\Big(\frac{1}{\ep_\mu}+1\Big)R
\nn
\\
&&
\qquad \qquad
+\,\,\frac14\,R_{\la\si}\,\bigg[
\frac{1}{30\ep_\mu}
\,-\,
\Big( \frac{1}{30}
+ \frac13\,\frac{m^2}{\cx}
+ \frac{m^4}{\cx^2} \,+\,\dots \Big)
\ln\Big( \frac{\cx}{m^2}\Big) \bigg] R^{\la\si}
\nn
\\
&&
\qquad \qquad
+\,\,\,
\frac12\,
R \bigg[ \frac{1}{\ep_\mu}\,
\Big(\frac{\tilde{\xi}^2}{2} - \frac{1}{180}\Big)
\,-\,
\Big(\frac{\tilde{\xi}^2}{2}
- \frac{1}{180}
+ \frac{\big(18 \tilde{\xi}^2
+ 6\tilde{\xi} - 1\big)}{18}\frac{m^2}{\cx\,}
\nn \\ && \qquad \qquad
-\,\,\,
\frac{\big(3 \tilde{\xi}  - 1\big)^2}{9}
\frac{m^4}{\cx^2\,}
\,+\,\dots \Big)
\ln\Big( \frac{\cx}{m^2}\Big) \bigg]
\,R\,\bigg\}\,.
\label{nonlocCC-UV}
\eeq
Here the superscript $^{(UV,2)}$ serves to remember that the nonlocal
part of this formula is the UV piece, up to the second order in
$m^2/k^2$, of the full expression (\ref{final}).

The first observation about expression (\ref{nonlocCC-UV}) is that
leading $\mathcal{O}(m^0/k^0)$ logarithm terms in the coefficients
correspond to the divergences of the fourth-derivative terms in
(\ref{final}), as it has to be. The sub-leading logarithmic terms 
have contributions
$\mathcal{O}(m^2/k^2)$ and  $\mathcal{O}(m^4/k^4)$.
The last kind of terms is of special interest to us in order to reproduce
(\ref{nonlocCC}).
The non-local terms in  (\ref{nonlocCC-UV}) come from the
fourth-derivative form factors in the general expression (\ref{final})
and are not connected with the renormalization of the CC term.

The reduction of the nonlocal $\mathcal{O}(m^4/k^4)$-terms
from expression (\ref{nonlocCC-UV}) to formula
(\ref{nonlocCCmass}), which we reproduced from
Ref.~\cite{Donoghue_22}, requires one more
operation. In the arguments of the logarithms in (\ref{Ops})
and (\ref{nonlocCC-UV}), one has to perform an \textit{ad hoc}
change, replacing $k^2 \to k^2+m^2$ and $\cx \to \cx+m^2$,
respectively. In principle, this is a legitimate operation
because the whole expression (\ref{nonlocCC-UV}) is an expansion
valid only in the UV regime $m^2 \ll k^2$ and making
such a replacement does not modify the coefficient of the
logarithm terms.\footnote{Let us mention that this also provides a
qualitative and apparent similarity with the complete expression
(\ref{final}) even in the IR decoupling regime, as it was explained
in \cite{Donoghue_22}.}
After the described operation, the second order,
$\mathcal{O}(m^4/k^4)$, terms in the brackets of
(\ref{nonlocCC-UV}) produce the desired structures
$\big(m^4/\cx^2\big)\ln \big(1+ \cx/m^2\big)$ exactly like
in (\ref{nonlocCCmass}).
It is worth noting that $\ln \big(1+ \cx/m^2\big)$ is different from
the correct nonlocal form factor given by Eq.~(\ref{A}). There is
a qualitative similarity between the two expressions, but it is not
quantitative. For instance, using the simplified formula
$\ln \big(1+ \cx/m^2\big)$, one cannot reproduce the correct
coefficient in the Appelquist and Carazzone decoupling theorem
in QED \cite{AC}, as this requires the correct expression (\ref{A}).

Finally, to complete the comparison with formula (\ref{nonlocCCmass}),
let us consider the minimal theory with $\xi=0$ or, equivalently, with
$\tilde{\xi}=-1/6$.
In this way, preserving only the $\mathcal{O}(m^4/k^4)$ logarithmic
terms, we arrive at an analog of (\ref{nonlocCCmass}), extracted
from the full form factor (\ref{final}),
\beq
\mathcal{L}^{(UV,2,nl)}_{scal}\,=\,-\,
\frac{m^4}{4(4\pi)^2}
\bigg\{
\Big( \frac{1}{\cx}R_{\la\si}\Big)
\ln\Big( \frac{\cx}{m^2}\Big)
\,\Big( \frac{1}{\cx}R^{\la\si}\Big)
\,-\,\frac{1}{2} \Big( \frac{1}{\cx}R\Big)
\ln\Big( \frac{\cx}{m^2}\Big)
\Big( \frac{1}{\cx}R\Big)
\bigg\}.
\label{nonlocCC-WR}
\eeq
Here and in other similar formulas presented below, superscript
$^{(UV,2,nl)}$ means that this is the expression for the second order
in $m^2/k^2$ nonlocal terms, extracted from the UV limit of the
\textit{fourth derivative terms} in the complete formula (\ref{final}).

The apparent difference between the two expressions (up to the
replacement $\cx \to \cx + m^2$ discussed above)
(\ref{nonlocCCmass}) and (\ref{nonlocCC-WR})
is only in the value of coefficients, which we cannot explain
with a complete certainty.
Let us note that the the ratio between
$R_{\la\si}^2$ and $R^2$ terms in (\ref{nonlocCCmass}) may be
caused by the aforementioned ambiguity (gauge fixing vs ignoring
 ${\hat T}_{3,4,5}$) that is inevitable when taking into account
only the two tensor structures ${\hat T}_1$  and  ${\hat T}_2$
in (\ref{biliH-T}). This ambiguity does not exist if one takes into
account all five structures in (\ref{biliH-T}) and do not restrict
the external metric $h_{\mu\nu}$ by a gauge fixing condition. This
is the way calculation of diagrams was done in \cite{apco} and
independently checked by the heat kernel method and in the
subsequent works \cite{CodelloZanusso2013}. For this
reason, we believe that the formula (\ref{nonlocCC-UV}) and (as
a particular case)  formula (\ref{nonlocCC-WR}), derived as the
sub-leading expansions from the correct UV result, are correct.

The main point is that contributions (\ref{nonlocCC-WR}) do
\textit{not} correspond to the genuine Einstein-Hilbert $R$-term
or to the CC term. Clearly, they are some pieces of the
form factors $k_W$ and $k_R$ of the fourth-derivative terms.
This conclusion is proved by the analysis of the tensor structure
of the polarization operator \cite{apco}.
Let us stress again that this analysis is possible only taking
into account all five tensor structures in (\ref{biliH-T}). It looks
impossible to extract some information about the nonlocal
contributions to the CC term from the logarithmic part of the
fourth-derivative form factors, especially because those are
leading and sub-leading UV contributions, while the CC term
manifests itself in the IR.

It is instructive to provide results similar to (\ref{nonlocCC-WR})
in the cases of massive fermion and vector fields.
The expressions similar
to (\ref{nonlocCC-UV}) can be easily derived from
the form factors given in \cite{fervi}, but we do not include these
technical details here.
For the fermion field, we find
\beq
\mathcal{L}^{(UV,2,nl)}_{ferm}\,=\,
\frac{m^4}{2(4\pi)^2}
\bigg\{\Big( \frac{1}{\cx}R_{\la\si}\Big)
\ln\Big( \frac{\cx}{m^2}\Big)
\,\Big( \frac{1}{\cx}R^{\la\si}\Big)
\bigg\}.
\label{nonlocCC-ferm}
\eeq
It is interesting that, compared to the scalar field, in
the $R_{\la\si}^2$ term there is the change of sign typical for the
divergences of the $m^4$-type (however, the coefficient is $-2$
instead of $-4$). On the other hand, the $R^2$ term cancels in this
case. Thus, the similarity with the $m^4$-type divergences in
(\ref{final}) is  a pure coincidence.

Finally, for the massive vector (Proca) model, we have
\beq
\mathcal{L}^{(UV,2,nl)}_{Proca}\,=\,
\frac{3m^4}{4(4\pi)^2}
\bigg\{\Big( \frac{1}{\cx}R_{\la\si}\Big)
\ln\Big( \frac{\cx}{m^2}\Big)
\,\Big( \frac{1}{\cx}R^{\la\si}\Big)
\,-\,\frac{1}{2} \Big( \frac{1}{\cx}R\Big)
\ln\Big( \frac{\cx}{m^2}\Big)
\Big( \frac{1}{\cx}R\Big)
\bigg\}.
\label{nonlocCC-vec}
\eeq
Up to the overall factor, this expression coincides with that
for massive scalar field (\ref{nonlocCC-WR}). We leave to
the interested reader to check that this coincidence concerns
only the $\mathcal{O}(m^4)$ nonlocal term and does not take
place in the zero- and second order terms of the expansion
analogous to (\ref{nonlocCC-UV}).

Thus, physical interpretation
of the $\mathcal{O}(m^2/k^2)$ and
$\mathcal{O}(m^4/k^4)$ terms is subtle. In the UV regime (where
$m^2$ in arguments of logarithms becomes irrelevant), these terms
really have a global scaling property identical to that of the CC
term. Even if the tensor structure of these terms forces us
attributing them to the $C^2$ and $R^2$ sectors of the action,
the scaling properties are those of the lower-derivative terms. In
this respect, one may use these expansions as an alternative to the
anomaly-induced actions for the quantum massive fields
at high energy (see, e.g., \cite{PoImpo} and further references
therein).

It is known that the anomaly-induced action modified
for nonzero masses of quantum fields has interesting
cosmological applications \cite{PoImpo,StabInstab}. It would be
certainly very interesting (albeit certainly more difficult) to
explore the effects of masses by using the complete
nonlocal action (\ref{final}). It is worth noting that in
the recent work \cite{LeoMedeiros-22} one can find a useful
formalism for the massless logarithm form factors. One may
expect that this interesting method can be adapted for a theory
with nonzero masses.  In this case, terms (\ref{nonlocCC}) and
(\ref{nonlocCC-ferm}) will appear in the action as the second order
corrections, while the first order terms (which we derived but
not analysed here) are the leading corrections.

\section{Conclusions}
\label{sec.5}

We analysed non-locality of quantum corrections due to free massive
scalar, fermion, vector fields to the effective gravitational action.
Although the genuine cosmological term does not admit a non-local
form factor, the form factors in the $C^2$ and $R^2$ terms produce
contributions (\ref{nonlocCC-WR}) which at energies larger than the
corresponding particle mass look like an effective cosmological term.
However, there is an important difference, as these
non-local analogs of the cosmological constant are, in fact, the
second-order terms in the $m^2/k^2$ expansions of the fourth
derivative terms in the UV.
Certainly, it would be very interesting to determine the impact of these
and other similar low-mass terms on the evolution of the early Universe.
It is worth adding that the role of $f(1/\cx R)$ terms in nonlocal
cosmology was analysed in \cite{Deser1,Deser2}.
In any case, the $\mathcal{O}(m^4)$-terms in
the expression (\ref{nonlocCCmass}) do not mean the running
of the cosmological constant, but represent the subleading terms
in the momentum-subtraction running of coupling constants of
marginal operators $R^2$ and $C^2$.

Similarly, calculations performed in a de Sitter background should
be correctly interpreted as those which describe the running of the
fourth derivative (Gauss-Bonnet and $R^2$) terms, rather than the
scale-dependence of the cosmological term. It is interesting that
these terms do not produce higher derivative ghosts on a flat
background and play very important role in the Starobinsky
inflationary model \cite{star}. Thus, the de Sitter - based relations
such as (\ref{Wood-IR}) may be a useful alternative to the
anomaly-induced effective action, especially for the
phenomenologically interesting large values of the nonminimal
parameter $\xi$.

\section*{Acknowledgments}
The authors are grateful to Denis Dalmazi for informing
about the misprint in one of the formulas.
I.Sh. is partially supported by Conselho Nacional de Desenvolvimento
Cient\'{i}fico e Tecnol\'{o}gico - CNPq (Brazil) under the grant
303635/2018-5 and by Funda\c{c}\~{a}o de Amparo \`a Pesquisa de
Minas Gerais - FAPEMIG under the project PPM-00604-18.


\end{document}